\documentstyle[11pt]{article}
\font\tenbf=cmbx10
\font\tenrm=cmr10
\font\tenit=cmti10
\font\elevenbf=cmbx10 scaled\magstep 1
 1
 1

\newcommand {\ignore}[1]{}

\newcommand{\noi}{\noindent}
\newcommand{\bc}{\begin{center}}
\newcommand{\ec}{\end{center}}

\def\ifmath#1{\relax\ifmmode #1\else $#1$\fi}
\def\3quarter{{\textstyle{3 \over 4}}}

\def\ra{\rightarrow}

\def\lf{\leaders\hbox to 1em{\hss.\hss}\hfill}

\def\21{$SU(2) \ot U(1)$}

\def\etal{\hbox{\it et al., }}

\def\eq#1{{eq. (\ref{#1})}}

\def\ltap{\raisebox{-.4ex}{\rlap{$\sim$}} \raisebox{.4ex}{$<$}}

\def\lsim{\raise0.3ex\hbox{$\;<$\kern-0.75em\raise-1.1ex\hbox{$\sim\;$}}}
\def\gsim{\raise0.3ex\hbox{$\;>$\kern-0.75em\raise-1.1ex\hbox{$\sim\;$}}}

\def\bel{\begin{letter}}
\def\eel{\end{letter}}
\def\beq{\begin{equation}}
\def\eeq{\end{equation}}
\def\bef{\begin{figure}}
\def\eef{\end{figure}}
\def\bet{\begin{table}}
\def\eet{\end{table}}
\def\bea{\begin{eqnarray}}
\def\ba{\begin{array}}
\def\ea{\end{array}}
\def\bi{\begin{itemize}}
\def\ei{\end{itemize}}
\def\ben{\begin{enumerate}}
\def\een{\end{enumerate}}
\def\ra{\rightarrow}
\def\ot{\otimes}

\def\eea{\end{eqnarray}}
\def\np#1#2#3{           {\it Nucl. Phys. }{\bf #1} (19#2) #3}
\def\pl#1#2#3{           {\it Phys. Lett. }{\bf #1} (19#2) #3}
\def\pr#1#2#3{           {\it Phys. Rev. }{\bf #1} (19#2) #3}
\def\prep#1#2#3{         {\it Phys. Rep. }{\bf #1} (19#2) #3}

\def\n.c.#1#2#3{         {\it Nuovo Cim. }{\bf #1} (19#2) #3}
\def\r.n.c.#1#2#3{       {\it Riv. del Nuovo Cim. }{\bf #1} (19#2) #3}

\relax
\begin{document}
\begin{center}{{\tenbf
               IMPLICATIONS OF A LIGHT GLUINO\\}
\vglue .6cm
{\tenrm F. DE CAMPOS}
\footnote{E-mail CAMPOSC at vm.ci.uv.es or 16444::CAMPOSC}\\
{\tenrm and}
{\tenrm JOS\'E W. F. VALLE}
\footnote{E-mail VALLE at vm.ci.uv.es or 16444::VALLE}\\
\baselineskip=13pt
{\tenit Instituto de F\'{\i}sica Corpuscular - C.S.I.C.\\
Departament de F\'{\i}sica Te\`orica, Universitat de Val\`encia\\}
\baselineskip=12pt
{\tenit 46100 Burjassot, Val\`encia, SPAIN         }\\
\vglue 0.3cm
{\tenrm ABSTRACT}}
\end{center}
\vglue 0.3cm
{\rightskip=3pc
 \leftskip=3pc
 \tenrm\baselineskip=11pt
 \noindent
We show that the possibility of having light gluinos
is not in conflict with recent LEP measurements
and also that cosmology does not rule
it out in any convincing way. In unified N=1
supergravity models, one expects that also the "photino" will be
light. Moreover it leads to $upper$ limits on the masses of
the other supersymmetric fermions, for example, the
lightest chargino should be lighter than about 75 GeV,
an thus detectable at LEP200.
\vglue 0.4cm}
\vglue 0.3cm
\hspace{\parindent}
There is a lot of controversy on whether or
not the existence of light gluinos, of mass
$2 \ltap m_{\tilde{g}} \ltap 6$ GeV, has been
confidently ruled out \cite{PDG92,UA1,Ant}. Although
this
possibility is theoretically marginal,
we have been recently encouraged by the fact
that light gluinos might account for the
apparent discrepancy between the values of
the strong coupling constant $\alpha_s$
as determined from low energy deep inelastic experiments
and those inferred from high energy LEP experiments.
The results of deep inelastic lepton-nucleon
scattering give $\alpha_s$ = 0.112 $\pm$ 0.005 at the
Z mass scale, which is lower than the results of $e^+ e^-$
analyses of event shapes. The averaged LEP value obtained
this way is {$\alpha_s$ = 0.124 $\pm$ 0.005} \cite{Bethke}.
Indeed, the evolution of $\alpha_s$
between 5 and 90 GeV is described much better if
we postulate a light gluino than if we do not
\cite{JezabekKuhn}. Similar arguments were also
given in ref. \cite{Clavelli} on the basis of
quarkonia data.

The existence of an electrically neutral coloured
fermion of relatively low mass could be envisaged as
a possible
solution which should not be overlooked unless one
can convincingly exclude it from experiment.
In supersymmetry the natural candidate is the
supersymmetric partner of
gluon, namely the gluinos, in the most popular
class of N=1 supergravity models
if gluinos are light one also expects
the "photino" to be light.
Considering a common supersymmetry breaking
mass parameter at the unification scale \cite{revsusy},
masses and mixing angles of the charged
and neutral supersymmetric fermions, charginos
and neutralinos, are determined by only
three independent parameters: the gluino mass,
which we may fix anywhere in the range of interest,
the ratio of the two Higgs vacuum
expectation values $\tan \beta = \frac{v_u}{v_d}$,
and the Higgsino mixing parameter $\mu$.

In this note we show that the possibility of
light gluinos in the context of supergravity
models is not in conflict with
LEP data. The relevant constraints may be
summarized as \cite{PDG92,lep}
\ben
\item
The limit on mass of the lightest of the charginos
$m_{\tilde{\chi}^{\pm}} \geq 45\mbox{ GeV}$
\item
The LEP limits on the total $Z$ width,
$\Gamma^{total}_{Z}=2.487\pm0.010\mbox{ GeV}$,
\item
The LEP limit on the invisible $Z$ decay width,
$\Gamma_{inv}=498\pm8\mbox{ MeV}$
\een
In addition to these we have also included
the LEP limit on the hadronic peak cross section
as well as the limits from $p\bar{p}$ colliders
on the ratio of W-to-Z cross sections
$0.825\leq\frac{R}{R_{SM}}\leq1.091$,
which could also be modified by the existence
of supersymmetric decay channels.

For fixed gluino mass one can determine
the region of supersymmetric parameters
allowed by the LEP experiments just in
terms of $\mu$ and $\tan\beta$
\footnote{We have not applied the constraint
$\tan\beta \geq 1$ that holds in models with
radiative electroweak breaking.}.  We have
determined what this region is for arbitrary
values of the gluino mass in the range
$ m_{\tilde{g}} \ltap 7$ GeV.
The result is shown in figure 1. We see that
there is a finite, but nonegligible butterfly-shaped
region of allowed $\mu$ and $\tan\beta$ values.
In particular, this shows explicitly that a very
light neutralino (most likely the LSP) is
perfectly consistent with the LEP measurements
of $\Gamma_Z^{invisible}$ if the supersymmetry
breaking gaugino mass parameters are small.
The reason for this is clear: in the limit
of strictly vanishing soft-breaking gaugino
masses the LSP is a pure photino, and it is
massless, forming an unbroken supersymmetric
multiplet with the photon. Such a state is
decoupled from the Z, and therefore no limits
can be set from LEP.

Now we move to cosmology.
If the LSP is a neutralino, almost pure photino,
then the relic LSP density will be, to a good
approximation, inversely proportional to the LSP
annihilation cross section which, in turn, is roughly
proportional to $m_{\tilde{\gamma}}^2/m_{\tilde{e}}^4$.
Requiring it not to be too large one obtains a lower
bound on the LSP mass, for this case:
$
m_{\tilde{\gamma}} \gsim 1 \rm{GeV} \times
({\frac{m_{\tilde{e}}}{45 \mbox{GeV}}})^2
\label{45}
$
which, from the LEP bound on selectrons, leads to
$m_{\tilde{\gamma}} \gsim 1$ GeV. For such masses
the annihilation of relic photinos is sufficient
to dilute their number density to an acceptable level.
If $m_{\tilde{\gamma}} \lsim 1$ GeV
one would have to rely on some photino decay mechanism
in order to avoid a conflict with the standard cosmological
picture. If R parity \cite{RP} is conserved, the photino will be
absolutely stable. Thus the simplest way out is to allow
for a small amount of R parity violation. Such possibility
is definitely allowed by experiment and there are several
extensions of the minimal supersymmetric standard model
where R parity violation can occur \cite{RPP,MASI,NPBRP}.
For example, in the case of spontaneous R-parity breaking
\cite{MASI} the photino decays mostly by majoron emission,
$\tilde{\chi}^0 \ra \nu + majoron$,a decay mode
basically unconstrained by astrophysics and
cosmology. On the other hand, the laboratory
missing energy signatures associated to the LSP
indistinguishable
from those it has in the minimal supersymmetric
standard model, to the extent that the invisible
decay is dominant.
We conclude that cosmological arguments can not
convincingly rule out the existence of a light photino.

We now note that in this class of supergravity models
it is possible, although marginally, to induce small
gluino and photino masses just as a result of radiative
corrections \cite{Masi} around 2 GeV,
mostly from a top-stop loop.

A most striking consequence of the light gluino
supergravity scenario is that the expected
mass spectrum in our class of light-gluino supergravity
models is characterized by relatively light supersymmetric
fermions. This follows from the allowed region of $\mu$
and $\beta$ values shown in figure 1. Indeed, for such
allowed parameter values the charginos
and neutralinos should be accessible at future accelerators
like LEP200.
For example, figure 2 shows the region of allowed chargino
masses. Clearly, one sees that the lightest chargino
should be lighter than about 75 GeV. A similar upper
bound also applies to the next-to-lightest of the neutralinos.
Moreover, if the relic photino population disappears only due to
selectron-mediated annihilations, also selectrons would
very close to the present limit, from \eq{45}. A quick
inspection at the renormalization group equations then
shows that the other sfermions would also be light in
this case. Although squarks are somewhat heavier, due
to colour, they too should lie in the region of sensitivity
of hadron collider experiments. One way to make the model
"safer" from being experimentally disproved would be to
allow for some R parity violation, so that the photino
can decay, as described above. This would relax the
limits on the sfermions masses, allowing them to
be heavier. However, the implied upper limits on
the chargino and neutralino masses would still hold.

In conclusion we would like to stress that
neither the existing data from LEP nor
cosmological considerations preclude the
possibility that a light photino exists,
as would be expected in N=1 supergravity
models where gluinos are sufficiently light
as to play a significant role in the running
of $\alpha_s$ between 5 and 90 GeV. The resolution of the
controversy on whether or not the existence
of light gluinos can be confidently ruled out
must rest upon the results of hadron colliders
and depend on the gluino lifetime and on the
details of the strongly interacting supersymmetric
spectrum. As suggested in \cite{JezabekKuhn},
future searches for evidence of a light gluino
in 4-jet  $e^+ e^-$ or 3-jet ep events should be pursued
by the experiments at LEP and HERA. Here we have
stressed the important complementary role played
by future searches at LEP200 for the electroweakly
interacting supersymmetric fermions. For example,
we have showed that the lightest chargino should
be lighter than about 75 GeV, with a similar
upper limit applying also to the neutralino
immediately heavier than the photino. These particles
should be accessible at LEP200.
\vskip .3cm
\vglue 0.2cm
{\elevenbf\noindent 6. Acknowledgements}
\vglue 0.2cm
\hspace{\parindent}
We thank M. Drees, G. Farrar and A. Masiero for stimulating
discussions and M. Gonzalez-Garcia for help with the
programing. This work was supported by the Spanish
Ministry of Education and Science under grant N. PB92-0084 (Spain)
and by a CNPq fellowship (F. de Campos)(Brazil).
\vskip .3cm
\noi
{\bf Figure Captions}\\
{\bf Fig 1}:\\
Regions of allowed $\mu$ and $\tan\beta$ values
in unified N=1 supergravity models with light gluinos.\\
{\bf Fig 2}:\\
Region of allowed chargino masses
in light-gluino supergravity models.
\bibliographystyle{ansrt}

\end{document}